# Quantum-Level Crosstalk Characterization of a 16x16 MEMS Optical Switch for Dynamic Quantum Communications

Persefoni Konteli
National and Kapodistrian University of Athens, Athens, Greece
perskonteli@di.uoa.gr

Nikolas Makris
National and Kapodistrian University of Athens, Athens, Greece
nikmakris@di.uoa.gr

Grigoris Anastasiou
National and Kapodistrian University of Athens, Athens, Greece
ganastasiou@di.uoa.gr

Konstantinos Tsimvrakidis
National and Kapodistrian University of Athens, Athens, Greece
kontsim@di.uoa.gr

George T. Kanellos
National and Kapodistrian University of Athens, Athens, Greece
gtkanellos@di.uoa.gr

***Abstract:*** *We present an all-to-all quantum-level crosstalk characterization of a commercial 16x16 MEMS optical switch for multi-user quantum communications using SNSPDs, correlating experimental data with a theoretical impact analysis on the decoy-state BB84 QKD protocol.*

## I. Introduction

The scope of quantum communication applications is rapidly expanding, from quantum computer interconnects to Quantum Key Distribution (QKD) and distributed quantum sensing. Because these technologies primarily rely on single-photon signals, they are highly susceptible to noise and often require stringent operational requirements. To achieve widespread, cost-efficient deployment, it is essential to integrate commercial photonic components, such as Dense and Coarse Wavelength Division Multiplexers (DWDM/CWDM) and Micro-Electro-Mechanical Systems (MEMS) optical switches [1], into quantum networks. The reliance on off-the-shelf optical components is especially critical in dynamic QKD networks [2]-[5], where every legitimate partner can communicate with any other legitimate partner, and in hybrid networks where high-speed classical optical channels coexist with QKD [6],[7]. This architecture is also highly applicable to data center topologies, where an optical switch is utilized to manage both quantum channels and classical signals acting as optical interconnects across diverse applications [8]. To realize such networks, cross-connect low-loss optical switches (LLOS) with excellent isolation are necessary to suppress crosstalk between the classical and quantum channels. Proper isolation minimizes noise and protects the vulnerable single photon detectors of the QKD modules from optical damage. While prior literature has explored device characterization, such as the Coherent-State Quantum Process Tomography (csQPT) proposed in [9], and the ν- optical time-domain reflectometry (OTDR) crosstalk analysis in [10], the impact of classical components on quantum network limits requires further investigation.

In this work, we present an in-depth analysis of the crosstalk induced by classical optical components that may limit the performance of quantum networks. Utilizing superconducting nanowire single photon detectors (SNSPDs), we systematically measure the photon leakage across every input to output port of a 16x16 MEMS LLOS. Our results indicate that the leakage for most port configurations does not significantly compromise the Secret Key Rate (SKR) of the QKD decoy-state BB84 protocol, especially at lower attenuation levels. Nonetheless, certain port combinations exhibit severe photon leakage (up to $2{\times}10^6$ cps). Our findings highlight that active switching operations must be carefully managed to avoid substantial performance degradation in the quantum channel.

## II. Experimental Setup

The experimental testbed is comprised by a C-band Continuous Wave (CW) tunable laser source emitting at 1550.12 nm (ITU channel grid C34) at a power of -10 dBm and a DiCon 16x16 optical MEMS LLOS. The LLOS provides a crosstalk isolation of over 70 dB and less than 0.8 dB insertion loss. On the detection side, four SNSPDs by Single Quantum were employed, exhibiting low dark count rate (DCR) down to 10 cps and excellent timing properties (11 ps jitter). The SNSPDs used are polarization sensitive requiring polarization compensation, as any sudden polarization drift can end up in reduced detection efficiency. To diminish the need for manual polarization compensation between the switching operations, four different SNSPDs were used to measure effectively the four different linear polarization bases (Horizontal, Vertical, Diagonal, Antidiagonal). To achieve the partition of the four polarization bases, a polarization analysis module (PAM) was employed while each output of the PAM was attached to a polarization controller (PC) keeping the polarization stable at each basis, as displayed in Fig. 1.

The experimental characterization was conducted in two phases, the automated switching of the classical laser signal across the LLOS, followed by the detection of photon leakage utilizing the SNSPDs. For each iteration, the classical optical signal was coupled into a single input port of the LLOS, while the SNSPDs monitored a specific output port. The classical signal was then sequentially swept across all output ports except the one linked to the SNSPDs in order to protect it from high power damage, detector saturation, or latching. This setup enabled the precise detection of single-photon leakage for each LLOS routing configuration of the classical signal. The entire procedure was systematically repeated across all input ports to completely map the leakage at every corresponding output. By analyzing the photon counting events recorded by the SNSPDs in a time-tagging module, we quantified the crosstalk of the LLOS, thereby providing a reliable estimation of how this noise translates to performance degradation in a QKD system.

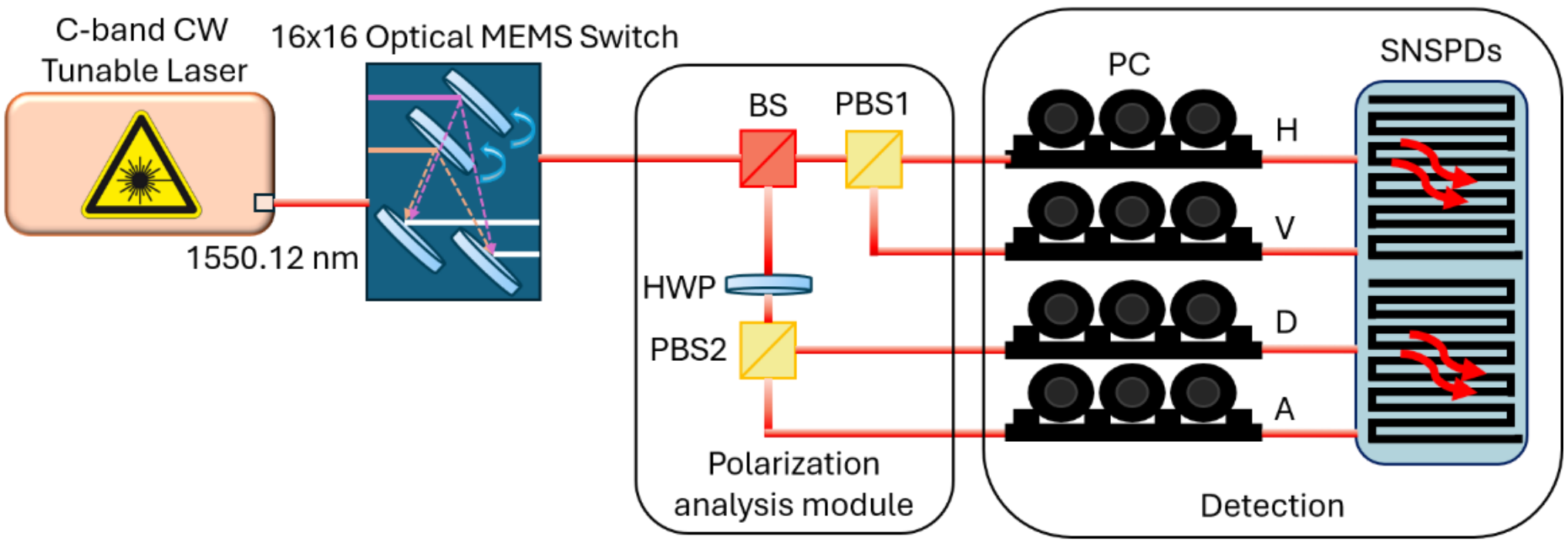

Fig. 1. The LLOS quantum-level crosstalk characterization setup.

## III. Results

During the experiment, the photon counts of all polarization bases were summed up for each input-output port combination, and then the mean value of ten iterations was calculated to measure the total photon count leakage in every different port connected to the SNSPDs. We observed that each output port exhibits a different crosstalk profile by plotting the photon counts for all the input-output combinations, as shown in Fig. 2. However, it is worth noticing that crosstalk is presented in specific patterns for all the output ports. These four patterns are the following; one output port (16) had minimum crosstalk for every case (<2500 cps), five output ports exhibited very high crosstalk only for a specific input or output port of the classical signal ($<4\times10^5$ cps), five output ports had low crosstalk for many input-output port combinations ($<25\times10^3$ cps), as displayed in Fig. 2a, b and c respectively, and in the other cases the crosstalk was high only for a few input-output port combinations, reaching up to $2\times10^6$ cps. These findings reveal an inconsistency on the crosstalk from port to port, since certain configurations presented higher crosstalk compared to others. This behavior can be attributed to the internal cross-connections of the LLOS and the overall architecture of the micromirrors in the switch arrays.

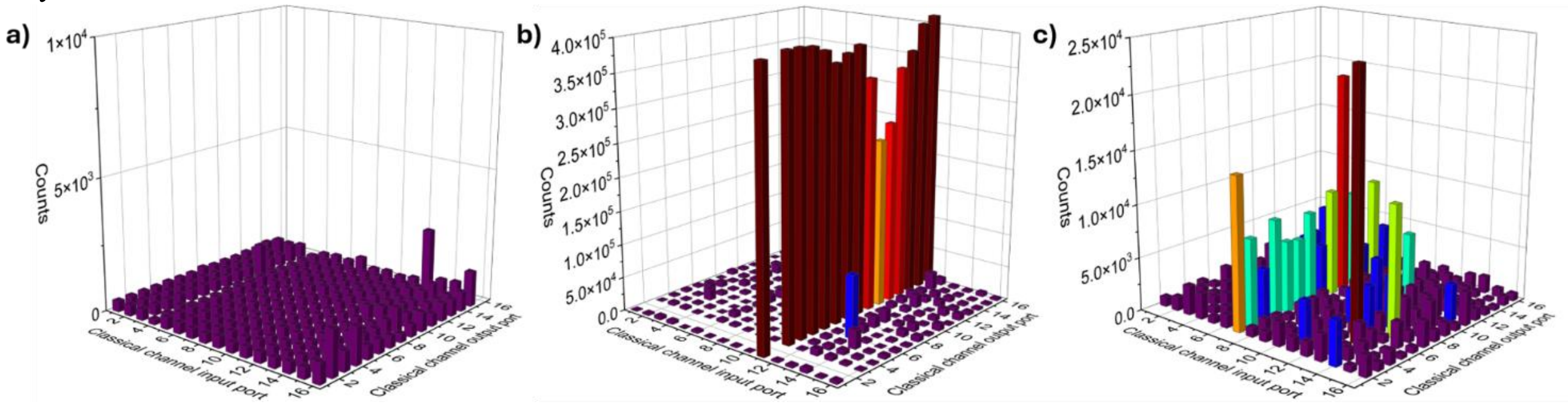

Fig. 2: Noise counts obtained for all classical input–output port combinations of a single quantum channel. The x- and y-axes represent the classical output and input ports, respectively, while the z-axis indicates the corresponding photon noise counts. Figures (a)–(c) illustrate different observed patterns, highlighting the non-uniform distribution of crosstalk-induced noise across the MEMS switch.

In Fig. 3, the empirical cumulative distribution function (ECDF) is displayed for the percentage of the input-output combinations as function of the crosstalk, which is considered as noise photons for the QKD systems. Additional stray noise, originating from external classical sources, can affect the quantum channel mainly in terms of achievable distance (or available link budget) and performance, expressed as the SKR. The impact differs depending on the employed protocol family and the specific protocol. Generally, both entanglement-based and prepare-and-measure (PM) protocols are robust to noise under low-loss conditions due to the high signal-to-noise ratio (SNR), but their performance and achievable distance are severely affected in high-loss conditions when noise is present [11],[12]. provides an example of the amount of noise required to cause a specific performance degradation. The results are obtained for the decoy-state BB84 protocol with a source repetition rate of 1 GHz.

TABLE I: Simulated noise thresholds corresponding to SKR performance degradation levels (-3dB, -10dB, and -20dB) under various baseline channel attenuation for the BB84 with decoy states framework [11].

| Link Loss | Noise/ -3db performance loss | Noise/ -10dB performance loss | Noise/ -20dB performance loss |
|---|---|---|---|
| 0 dB | 804321 | 1543572 | 1720732 |
| 10 dB | 78271 | 149921 | 167061 |
| 20 dB | 7811 | 14951 | 16661 |

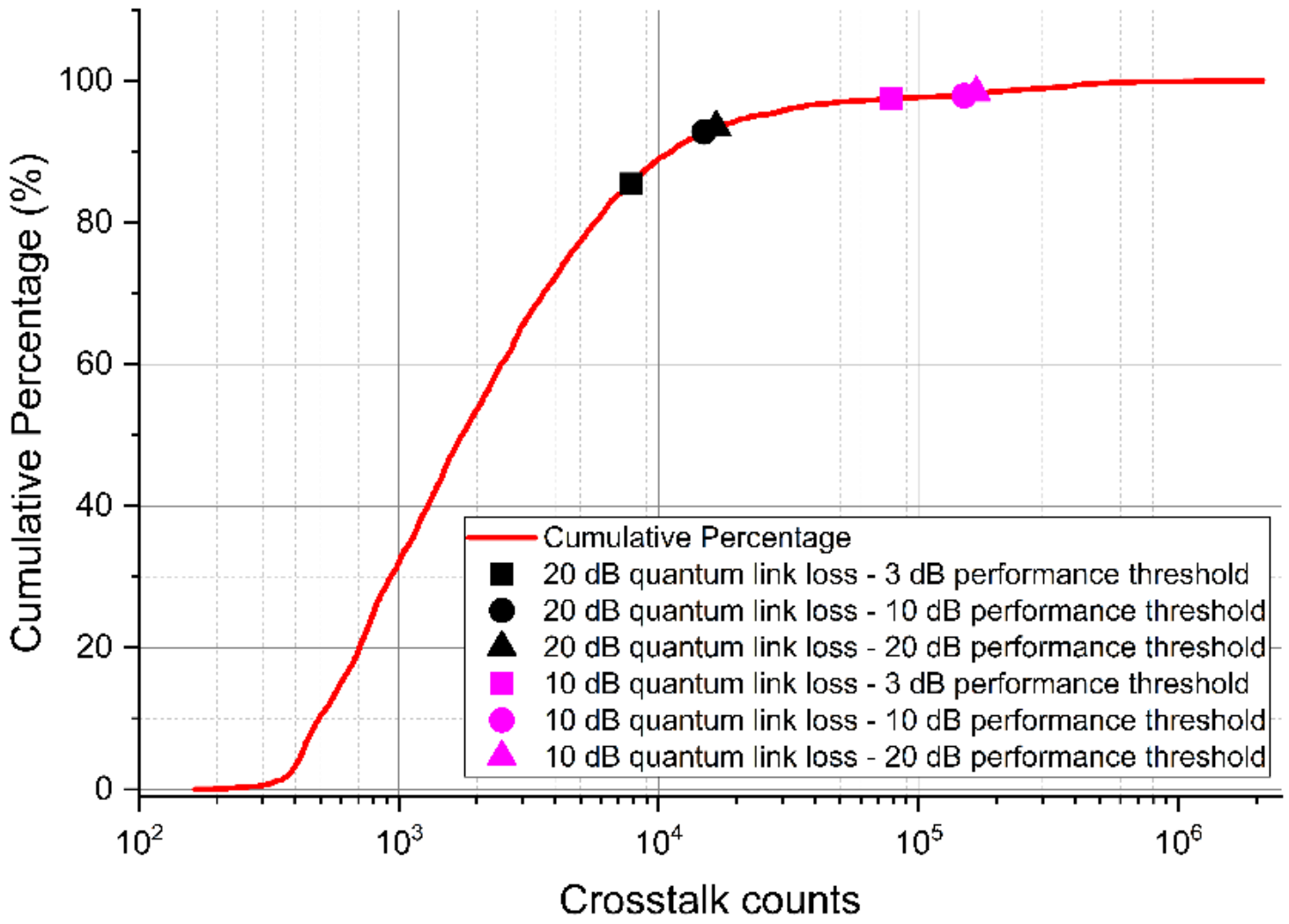

Fig. 3. ECDF of the measured crosstalk noise counts across all input–output port combinations. The curve represents the cumulative percentage of port combinations with noise counts below a given value. The markers denote simulated noise thresholds corresponding to SKR degradation levels of −3 dB, −10 dB, and −20 dB for baseline quantum channel losses of 10 dB (magenta color symbols) and 20 dB (black color symbols).

It is evident that under low-loss conditions, the amount of noise required to cause a performance deterioration is higher, whereas under higher-loss conditions this threshold is reduced as the SNR decreases and the relative contribution of noise becomes more significant. In Fig. 3, the scatter symbols represent threshold values obtained from the theoretical analysis presented in TABLE I, that help identify the percentage of port combinations for which performance drop would occur. Specifically, for 20 dB attenuation of the quantum channel, 14.53%, 7.21%, and 6.51% of the port combinations would cause performance losses of 3 dB, 10 dB, and 20 dB, respectively. For 10 dB quantum channel attenuation, the corresponding values are 2.50%, 2.09%, and 1.67%. For the 0 dB attenuation scenario, only four port combinations would cause significant performance deterioration by exhibiting excessive leakage. As expected, configurations with lower loss for the quantum link exhibit higher SNR and thus higher resistance to performance degradation due to noise. Consequently, most of the input-output port configurations enable quantum communication since they exhibit a low noise profile, yet switching operations should be carefully executed based on the crosstalk characterization of each LLOS considering that specific ports can lead to performance deterioration or blinding of the detectors.

## IV. Conclusions And Discussion

We experimentally carried out a quantum-level crosstalk characterization of a 16x16 MEMS LLOS using SNSPDs, motivated by the noise sensitivity of quantum communication technologies, with a specific focus on QKD applications. Furthermore, we conducted a theoretical analysis to evaluate the impact of the crosstalk-induced photon noise on the widely used decoy-state BB84 QKD protocol. Our simulations indicate that while a few specific port configurations exhibit high noise levels capable to severely degrade the system's performance and dynamic range, the vast majority of combinations are fully compatible with quantum communications, supporting link budgets up to 20 dB. Notably, the ability of the QKD protocol to maintain a sufficient SKR under 20dB of attenuation, confirms that integrating commercial LLOSs into dynamic QKD networks is viable. Additionally, a critical aspect is the non-uniform behavior of the noise figures, which may derive from the characteristics of the LLOS itself, such as the design of the micromirrors and the beam-steering-based elements or other physical attributes of the components. Thus, deploying commercial off-the-shelf LLOSs in dynamic QKD networks is a highly practical and cost-effective alternative that potentially precludes the need for specialized quantum optical components. Future studies must focus on an extensive characterization of diverse LLOS architectures to experimentally validate the peculiarities of individual switches under various QKD protocols.

## Acknowledgments

This work was funded by the EU project ORQESTRA - Quantum technologies (GA 101224573).